\documentclass[]{emulateapj}
\shorttitle{Luminosity functions of LMXBs in Centaurus A}
\shortauthors{Voss et al.}

\begin{document}

%% LaTeX will automatically break titles if they run longer than
%% one line. However, you may use \\ to force a line break if
%% you desire.

\title{Luminosity functions of LMXBs in Centaurus A:\\
globular clusters versus the field}

%% Use \author, \affil, and the \and command to format
%% author and affiliation information.
%% Note that \email has replaced the old \authoremail command
%% from AASTeX v4.0. You can use \email to mark an email address
%% anywhere in the paper, not just in the front matter.
%% As in the title, use \\ to force line breaks.

\author{Rasmus Voss\altaffilmark{1,2}}
\email{rvoss@mpe.mpg.de}

\author{Marat Gilfanov\altaffilmark{3,4}}

\author{Gregory R. Sivakoff\altaffilmark{5,8}}

\author{Ralph P. Kraft\altaffilmark{6}}

\author{Andr\'es Jord\'an\altaffilmark{6,7}}

\author{Somak Raychaudhury\altaffilmark{9}}

\author{Mark Birkinshaw\altaffilmark{10}}

\author{Nicola J. Brassington\altaffilmark{6}}

\author{Judith H. Croston\altaffilmark{11}}

\author{Daniel A. Evans\altaffilmark{12,6}}

\author{William R. Forman\altaffilmark{13}}

\author{Martin J. Hardcastle\altaffilmark{11}}

\author{William E. Harris\altaffilmark{14}}

\author{Christine Jones\altaffilmark{13}}

\author{Adrienne M. Juett\altaffilmark{15}}

\author{Stephen S. Murray\altaffilmark{6}}

\author{Craig L. Sarazin\altaffilmark{5}}

\author{Kristin A. Woodley\altaffilmark{16}}

\author{Diana M. Worrall\altaffilmark{10}}

%% Notice that each of these authors has alternate affiliations, which
%% are identified by the \altaffilmark after each name.  Specify alternate
%% affiliation information with \altaffiltext, with one command per each
%% affiliation.

\altaffiltext{1}{Max-Planck-Institut f\"ur extraterrestrische Physik, 
Giessenbachstrasse, D-85748, Garching, Germany}
\altaffiltext{2}{Excellence Cluster Universe, Technische Universität M\"unchen, Boltzmannstr.
2, D-85748, Garching, Germany}
\altaffiltext{3}{Max Planck Institut f\"ur Astrophysik, Karl-Schwarzschild-Str. 1, D-85741, Garching, Germany}
\altaffiltext{4}{Space Research Institute, Russian Academy of Sciences, Profsoyuznaya
84/32, 117997 Moscow, Russia}
\altaffiltext{5}{Department of Astronomy, University of Virginia, P.O. Box 400325,
Charlottesville, VA 22904-4325}
\altaffiltext{6}{Harvard/Smithsonian Center for Astrophysics, 60 Garden Street, MS-67, Cambridge, MA 02138}
\altaffiltext{7}{Departamento de Astronom\'{\i}a y Astrof\'{\i}sica, Pontificia Universidad Cat\'olica
de Chile, Casilla 306, Santiago 22, Chile}
\altaffiltext{8}{Department of Astronomy, 4055 McPherson Laboratory, Ohio State 
University, 140 West 18th Avenue, Columbus, OH 43210-1173}

\altaffiltext{9}{School of Physics and Astronomy, University of Birmingham, Edgbaston, Birmingham B15 2TT, UK}
\altaffiltext{10}{Department of Physics, University of Bristol, Tyndall Avenue, Bristol BS8 1TL, UK}
\altaffiltext{11}{School of Physics, Astronomy and Mathematics, University of Hertfordshire, College Lane, Hatfield, Hertfordshire AL10 9AB, UK}
\altaffiltext{12}{MIT Kavli Institute, 77 Massachusetts Avenue, Cambridge, MA 02139}
\altaffiltext{13}{Smithsonian Astrophysical Observatory, Harvard-Smithsonian Center for Astrophysics, Cambridge, MA 02138}
\altaffiltext{14}{Department of Physics and Astronomy, McMaster University, Hamilton, ON L8S 4M1}
\altaffiltext{15}{NASA Postdoctoral Fellow, Laboratory for X-ray Astrophysics, NASA Goddard Space Flight Center, Greenbelt, MD 20771}
\altaffiltext{16}{Department of Physics and Astronomy, McMaster University, Hamilton, ON L8S 4M1}

%% Mark off your abstract in the ``abstract'' environment. In the manuscript
%% style, abstract will output a Received/Accepted line after the
%% title and affiliation information. No date will appear since the author
%% does not have this information. The dates will be filled in by the
%% editorial office after submission.
\newpage
\begin{abstract}
We study the X-ray luminosity function (XLF) of low mass X-ray binaries (LMXB)
in the nearby early-type galaxy Centaurus A, concentrating primarily on two aspects of binary populations: the 
XLF behavior at the low luminosity limit and comparison between globular cluster and field sources. 
The 800 ksec exposure of the deep \textit{Chandra} VLP program allows us to reach a limiting luminosity of
$\sim 8\cdot 10^{35}$ erg s$^{-1}$, about $\sim 2-3$ times deeper than previous investigations.
We confirm the presence of the low luminosity break in the overall LMXB XLF at $\log(L_X)\approx 37.2-37.6$ below 
which the luminosity distribution  follows a $dN/d(\ln L)\sim const$ law. Separating globular cluster and field sources, 
we find a statistically significant difference between the two luminosity distributions with a relative 
underabundance of faint sources in the globular cluster population. 
This demonstrates that the samples are drawn from distinct parent populations and may disprove the hypothesis that the entire LMXB population in early type galaxies is created dynamically in globular clusters.
As  a plausible explanation for this difference in the XLFs, we suggest that there is an enhanced fraction of  
helium accreting systems in globular clusters, which  are created in collisions between red giants and
neutron stars. Due to the 4 times higher ionization temperature of He, such systems are subject to accretion disk instabilities at $\approx 20$ times higher mass accretion rate, and therefore are not observed as persistent sources
at low luminosities. 
\end{abstract}

%% Keywords should appear after the \end{abstract} command. The uncommented
%% example has been keyed in ApJ style. See the instructions to authors
%% for the journal to which you are submitting your paper to determine
%% what keyword punctuation is appropriate.


%% From the front matter, we move on to the body of the paper.
%% In the first two sections, notice the use of the natbib \citep
%% and \citet commands to identify citations.  The citations are
%% tied to the reference list via symbolic KEYs. The KEY corresponds
%% to the KEY in the \bibitem in the reference list below. We have
%% chosen the first three characters of the first author's name plus
%% the last two numeral of the year of publication as our KEY for
%% each reference.

%% Authors who wish to have the most important objects in their paper
%% linked in the electronic edition to a data center may do so by tagging
%% their objects with \objectname{} or \object{}.  Each macro takes the
%% object name as its required argument. The optional, square-bracket 
%% argument should be used in cases where the data center identification
%% differs from what is to be printed in the paper.  The text appearing 
%% in curly braces is what will appear in print in the published paper. 
%% If the object name is recognized by the data centers, it will be linked
%% in the electronic edition to the object data available at the data centers  
%%
%% Note that for sources with brackets in their names, e.g. [WEG2004] 14h-090,
%% the brackets must be escaped with backslashes when used in the first
%% square-bracket argument, for instance, \object[\[WEG2004\] 14h-090]{90}).
%%  Otherwise, LaTeX will issue an error. 

\section{Introduction}
With the advent of \textit{Chandra}, the study of populations of X-ray
binaries in nearby galaxies became possible 
\citep[e.g.][]{Sarazin2000,Kraft2001,Kundu2002}. In young
stellar populations the high mass X-ray binaries dominate, and
the luminosity function (LF) was found to be a simple power law $dN/dL_X\propto L_X^{-\Gamma}$ 
with a differential slope of
$\Gamma\sim1.6$ \citep{Grimm}, with an indication of flattening at very 
low luminosities $\lesssim\log(L_X)=35.5$, probably due to the propeller 
effect\footnote{At low mass accretion rates the centrifugal barrier imposed by the  
magnetosphere of the spinning neutron star may inhibit the flow of  
matter towards the neutron star, thus quenching the X-ray emission \citep{Illiaronov}.} \citep{Shtykovskiy}.
The LF in old stellar populations is dominated by low mass X-ray
binaries (LMXBs) and was shown to be steep
at the bright end, $\log(L_X)>37.5$, with a power law index in the
$\sim1.8-2.5$ range \citep{Gilfanov,Kim}, flattening to $dN/dL\propto L^{-1}$ below $\log(L_X)\lesssim37.5$
\citep{Gilfanov,Voss-m31}. 
While the low luminosity break is observed in the bulges of spiral galaxies, there is no consensus on
the shape below $\sim10^{37}$ erg s$^{-1}$ in elliptical galaxies
\citep{Voss-cena,Kim2006}. Currently the only elliptical galaxy in
which it is possible to observe X-ray sources well below $10^{37}$ erg s$^{-1}$ is 
Centaurus~A (Cen~A).

The number of LMXBs per unit stellar mass is known to be $\sim$
two orders of magnitude higher in Galactic globular clusters (GCs) than in
the field of the Milky Way \citep{Clark}. Also in external galaxies, the 
frequency of LMXBs is particularly high in GCs \citep[e.g.][]{Sarazin,Minniti2004,Jordan2007,Posson2009}, and this is attributed to dynamical processes
in which LMXBs are formed in close stellar encounters.
The same mechanisms also have
been shown to be responsible for the formation of a significant number
of "surplus" LMXBs in the dense inner bulge of M~31 \citep{Voss-coll}.
It is currently a subject of debate whether the entire LMXB population in (early type) galaxies was formed in GCs 
\citep{White,Kundu2002,Maccarone,Juett,Irwin,Humphrey,Kundu2007}.
This suggestion has been confronted by the statistics of LMXBs and globular clusters in spiral 
galaxies and especially in the Milky Way.  
Also, evidence has been found that the luminosity distributions of LMXBs in globular clusters (and in 
the central parts of M~31) and in the field differ at the low luminosity end, below $\sim10^{37}$ erg s$^{-1}$,  
\citep{Voss-m31,Fabbiano2007,Woodley2008,Kim2009}, which would be a strong indication of different formation histories.

Cen~A is one of the best available candidates for the study of the LF of LMXBs
in an early-type galaxy, as it is massive enough to contain
a sufficient number of LMXBs and is sufficiently nearby
that a meaningful sensitivity can be achieved in a reasonable exposure time  
with \textit{Chandra}.
Previous studies of Cen A with \textit{Chandra} have yielded information
on the nucleus \citep{Evans}, the interstellar medium \citep{Kraft2003,Kraft2007,Kraft2008,Croston2009}, 
the jet \citep{Kraft2002,Hardcastle2003,Hardcastle2007,Worrall2008}, and the shell structures 
\citep{Karovska}.
The off-center population of point sources was first studied by
\citet{Kraft2001}. \citet{Minniti2004} investigated optical counterparts of
X-ray sources, and the luminosity function and spatial distribution
of the LMXBs were investigated by \citet{Voss-cena}.
\citet{Jordan} and \citet{Woodley2008} studied
the connection between GCs and LMXBs, and
\citet{Sivakoff} studied a transient black hole candidate.

Recently, Cen A has been the target of a \textit{Chandra} VLP program, which brought the total exposure time to $\sim$800 ks. With this exposure, individual point sources can be detected down to the luminosity of $\sim 6\cdot 10^{35}$ erg/sec and the population  of compact sources as a whole  can be studied in a statistically complete manner down to 
$\sim 8\cdot 10^{35}$ erg/s.  The significant, about 4-fold, increase in the exposure is advantageous for the study of the astrophysics of compact sources. The results of the detailed analysis of these data including the source lists will be published in forthcoming papers.
In this paper we take advantage of the increased \textit{Chandra} exposure
to perform a deeper study of the luminosity distribution of LMXBs,
with the main focus on the XLF behavior in the low luminosity regime and the difference between the LFs of the field and globular cluster LMXBs.

\begin{figure}
\resizebox{\hsize}{!}{\includegraphics[angle=0]{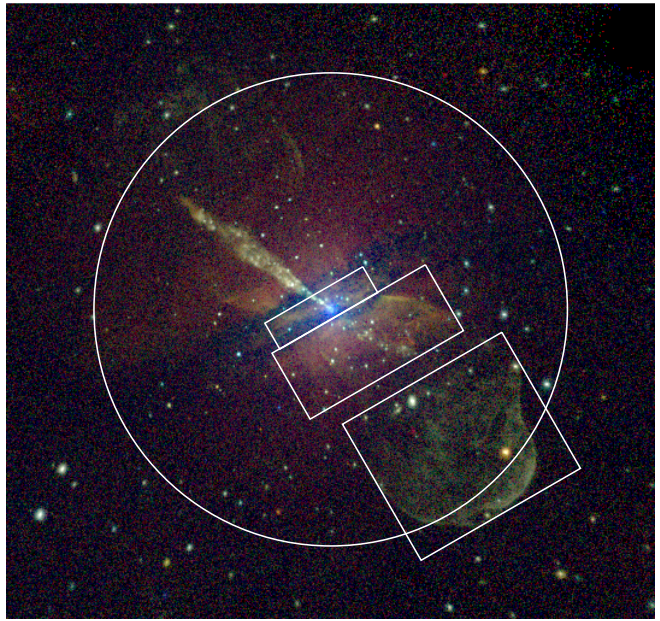}}
\caption{The combined X-ray 0.5-8 keV image of Cen A with pixel size 0.492$\times$0.492 arcsec$^{2}$. The region analysed in this paper is inside the circle with a 5 
arcmin radius. In the rectangular regions, only sources with a
luminosity above $5\times10^{36}$ erg s$^{-1}$ were used, and sources
in the jet were excluded. The image was prepared
using the data reduction described in the text, including exposure correction
and the removal of readout streaks. Besides the large number of X-ray point
sources, the jet and the south-west bubble and the absorption lanes 
are also clearly evident. Traces of a north-west bubble and a counter-jet can also be seen.}
\label{fig:nice}
\end{figure}

\section{Data Analysis}
\label{sect:analysis}
We combined all the available \textit{Chandra} ACIS observations of Cen A, to
obtain the deepest possible combined image. With the high number of point
sources, it was possible to align the individual images to a relative precision of
0.1 arcsec. The individual observations are listed in Table \ref{tab:obs}, 
and the data preparation will be described in
Kraft et al. (in prep). We found that for the point-source detection 
and luminosity estimation the removal of time intervals with high
background was not necessary,
as the increased exposure time outweighs the increased background.
In each observation readout streaks are seen, due
to the luminous central source.
To avoid the detection of spurious sources, the streaks were removed
from individual images before combining them. As the telescope roll angle
varies between the observations, the removed areas 
are well covered by other observations.
The following analyses were done on images filtered to retain only photons in
the energy range 0.5-8.0 keV.
An exposure map for the combined image was created assuming a power-law spectrum
with a photon index of $\Gamma=1.7$ (while the spectra of individual sources
vary, this model provides a good fit to the average spectrum of LMXBs in the
\textit{Chandra} band),
and Galactic foreground absorption of $8.4\times10^{20}$ cm$^{-2}$ \citep{Dickey}.
The exposure map gives the effective area of the observations at each pixel,
taking into account effects such as vignetting, the quantum efficiency degradation
of the detectors and the dithering of the telescope.  
We used CIAO {\tt wavdetect} to detect sources, using the same parameters as
in \citet{Voss-cena,Voss-m31}. Compared to the default settings, this
ensures that all relevant detection scales are sampled, and the
background cleaning is slightly improved.

For each source we estimate the point-spread-function (PSF)
by averaging the PSFs of the individual observations, obtained from
the \textit{Chandra} PSF library. These were weighted
by the values of the exposure maps at the source position, and assuming
a constant count rate for a source in all observations. While the sources
do vary, only a few sources are highly variable, and the effect of this on 
the luminosity function has been found to be small
\citep[e.g.][]{Voss-cena}. We compiled a list of all sources within 10 arcmin
from the center of Cen A, applying
a detection threshold of $10^{-6}$ (yielding
an average of 1 false source per $10^6$ 0.492$\times$0.492 arcsec$^{2}$ pixels), 
giving an expectation of approximately 5 spurious detections. In the analysis
below we use various subsamples optimized to the specific analyses (e.g.
sources in the 7.5-10 arcmin annulus to analyze the background contamination,
and sources within the 5 arcmin circle to analyze the LF of field LMXBs).
A second sample of sources was compiled, using a detection threshold
of $10^{-5}$, giving a higher sensitivity at the cost of 10 times more
spurious sources. 

This list was used for investigating sources coincident
with globular clusters (see below). Given the relatively small area covered
by each GCs in the image, 
only 0.25 spurious sources coincident with GCs are expected for this
threshold.
The count rate was determined
using circular apertures with an encircled energy of 85\% (radii varying
between 1 and 10 arcsec), correcting
for contamination from nearby sources, as described in \citet{Voss-m31}.
The luminosities of the point sources were then calculated, assuming a power-law
spectrum with $\Gamma=1.7$ and with Galactic foreground absorption, and a distance to Cen A of
3.7 Mpc \citep{Ferrarese}. \\

\subsection{Contamination by diffuse emission structures}
\label{sec:contamination}

\begin{figure}
\resizebox{\hsize}{!}{\includegraphics[angle=0]{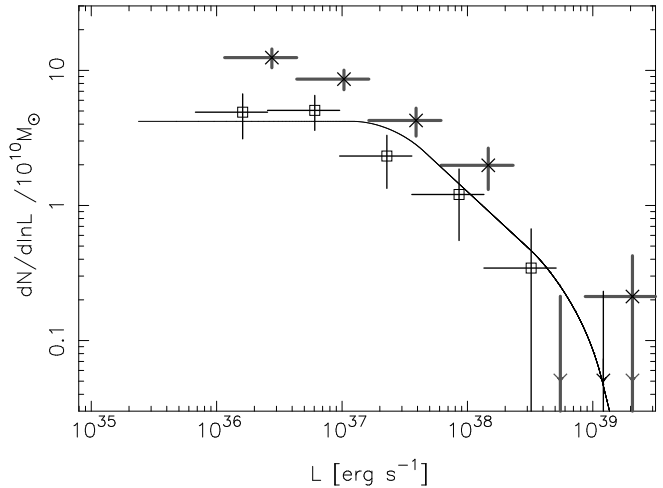}}
\caption{Contamination of the luminosity function by structures in the extended emission.
The luminosity functions of all compact sources detected by {\sc wavdetect} inside (crosses) and outside (open squares) problematic regions marked by boxes in Fig.\ref{fig:nice}. Only sources within the 5 arcmin circle in Fig.\ref{fig:nice} were used, and the jet and nucleus regions are excluded in both LFs. The contribution of background sources is subtracted, and LFs are corrected for incompleteness and normalized to the enclosed stellar mass, as discussed in section \ref{sec:lfs}. Above $5\times10^{36}$ erg s$^{-1}$ structures can be reliably identified and removed.}
\label{fig:lf_good_bad}
\end{figure}

For the analysis of the population of X-ray point sources, Cen A is in many
ways a complicated galaxy. It has a strongly warped dust disc with evidence
for star formation. This leads to heavy absorption, and therefore
uncertain luminosities, of the X-ray sources
in these lanes, as well as a possible contribution to the source sample from
high-mass X-ray binaries. The strong emission from the X-ray jet 
\citep{Kraft2002,Hardcastle2007,Worrall2008} and
its structure of many knots make it difficult to detect and identify point 
sources there. For this reason, all sources in the
jet region and within a radius of 20 arcsec from the nucleus were
excluded from the following analysis (see \cite{Voss-cena} for exact definition of these regions).
In the counter-jet direction, further away from the center
of the galaxy, there is a large number of filaments and shell structures
\citep{Karovska}. These features are all visible in Fig. \ref{fig:nice},
where the combined X-ray image of Cen A is shown.
Many of the features in these regions cannot be
clearly distinguished from faint point sources, leading to contamination
of the source sample, as illustrated by the Fig.\ref{fig:lf_good_bad}. The figure shows the LFs of all sources, detected by {\sc wavdetect} task, inside and outside the regions occupied by filaments and shell-like structures 
(including sources that are obviously part of diffuse structures that
are removed from the final sample, see below). The LFs are normalized to the stellar mass and show a higher specific frequency of compact "sources" in the problematic regions, with the contamination increasing severely towards the low luminosity end. 

We attempted to clean the point source list, inspecting the image visually and removing all obvious members of extended structures. We found that this procedure works reasonably well for bright sources but fails at the faint end of the luminosity distribution. On the other hand, the problematic regions contain a rather large fraction of the stellar mass, more than a half inside the 5 arcmin area. Given the rather limited number of bright sources in the galaxy, entirely excluding these regions from the analysis would affect notably the statistical quality of the bright end of the luminosity distribution. For this reason we chose to retain relatively bright sources from the problematic regions in the sample,  
for fluxes above $3.05\times10^{-15}$ erg s$^{-1}$ cm$^{-2}$
($>5\times10^{36}$ erg s$^{-1}$ assuming the distance of Cen A). Above this flux
it is possible to distinguish between diffuse and point-source emission. Sources that were evidently part of extended structures
were identified by eye and removed from the sample. After this removal, 
34 sources with luminosities above $5\times10^{36}$ erg s$^{-1}$ were left in the
source list from the contaminated regions. To verify the outcome of this procedure we compared the specific source frequency, per unit stellar mass, in broad luminosity bins, inside and outside the regions dominated by the extended structures and found a good agreement. 
While we are confident that the majority of false source
detections are eliminated by this procedure, a small residual 
contamination of the sample at the faint end of the luminosity function may still remain. 
As an additional check, we verified that all results reported below are reproducible with the source lists based on the "good" regions  only, albeit with larger errors and/or smaller statistical significance.

Finally, based on the previous X-ray catalogues of the Cen A point sources, we identified 5 sources
as foreground stars \citep[see source catalogue of][]{Voss-cena}, 
and they were also removed from our sample.

\subsection{Identification of globular cluster sources}

We identified GC counterparts using the catalogues of 
\citet{Woodley2007} and \citet{Jordan2007,Jordan}, and a search radius of 2 arcsec.
As described above, the X-ray point source list obtained
with a relaxed detection threshold of $10^{-5}$ was used for cross-correlation with the GC catalogs. 
This gave a total of  49 GC X-ray sources\footnote{The sum of the GC 
LMXBs in table \ref{tab:ratios}
only add up to 47, as two sources are detected below $10^{36}$ erg s$^{-1}$.} with an expectation of 0.25 false sources (see above) and of 
4.25 random coincidences, the latter estimated by displacing the X-ray source 
list by 5 arcsec in various directions.  We did not find additional
matches in the catalogues of \citet{Harris2004,Peng2004}. Finally we
did not use the catalogue of \citet{Minniti2004}. This catalogue was
compiled by searching for GCs at the position of X-ray point sources.
As the X-ray catalogue used in their study is significantly shallower 
than ours, it is biased towards bright sources, and such a bias would
be problematic for our luminosity function analysis.
Inside the search area of the globular clusters,
the expected number of background sources is $\lesssim1$, and therefore most
of the random coincidences are with field (non-GC) LMXBs.
The catalogue of \citet{Jordan}
covers 68\% of the area investigated in this paper.
It is based on \textit{Hubble Space Telescope} images, and is significantly
deeper than the catalogue of \citet{Woodley2007}, which covers the entire
area investigated in this paper. However, of the
40 GC X-ray sources detected inside the area included in the 
\textit{Hubble Space Telescope} catalogue, only 11 were
not found in the catalogue of \citet{Woodley2007}. 
Most of the GCs harboring X-ray sources ($\sim70\%$) were picked up by the 
shallower catalogue because the more luminous clusters are more
likely to host X-ray sources \citep[e.g.][]{Sarazin,Jordan2004,Sivakoff2007}. 
With 7 GCs hosting LMXBs observed outside
the area included in the \textit{Hubble Space Telescope} catalogue, we
therefore estimate that there
may be only $\sim3$ undetected GC  associated with X-ray sources in
our sample. This number is small enough to be neglected.

We note that
most of the areas unobserved by the \textit{Hubble Space Telescope} are in
regions of relatively low stellar and GC density (along the minor axis of Cen A)
and therefore the number of associations between GCs and LMXBs does not scale
with the area of the region. 
7 X-ray sources  match the GC candidates from the 
catalogue of \citet{Minniti2004}, but are not identified as GCs in the analysis of \citet{Jordan2007,Jordan}. In all these cases we gave preference to the classification based on  the \textit{Hubble Space Telescope} data.

\section{Luminosity functions}
\label{sec:lfs}

For the analysis of the overall LF of all LMXBs we use the source sample constructed as described in the section \ref{sec:contamination}. In order to limit the background-source contamination, we use only sources within 5 arcmin from the center of the galaxy.   
This area is indicated by
the circle in Fig. \ref{fig:nice}.
With the half-light radius
of Cen A being $\sim$5 arcmin \citep{Dufour},
the vast majority of field sources outside
this region are background sources \citep[see e.g. the radial source
distribution in Fig. 4 of][]{Voss-cena}. 
Moreover, the incompleteness effects
become severe, and the large point spread function 
makes it difficult to distinguish point-like and extended sources. 
The reduced quality of the LF due to these effects outweighs the small increase
in the number of LMXBs. 
We further construct two sub-samples: the field (non-GC) and the GC sources. The field sources are a selection of non-GC sources from our main sample. In building the sample of GC sources we used the entire galaxy, due to the much reduced probability of contamination by background sources and false sources from sub-structures in the extended emission.

With our selection
criteria, the regions  used to assemble the field source
sample cover about half of the stellar mass of the galaxy
at $L_X>5\cdot 10^{36}$ erg/s and $\sim25$\% at the faint end. 
In absolute units this corresponds to $6.8\times10^{10} M_{\odot}$ and  $3.3\times10^{10} M_{\odot}$ respectively.
These numbers were calculated  from the \textit{K}-band light in the \textit{2MASS} Large Galaxy
Atlas \citep{Jarret} using $M_{\ast}/L_K\simeq0.76$ \citep{Bell} appropriate for the
$(B-V)\simeq0.88$ colour of Cen A.
The field sample consists of 154 sources above 
a flux of $4.3\times10^{-16}$ erg s$^{-1}$ cm$^{-2}$ 
($\sim$the completeness limit, corresponding to a luminosity
of $7\cdot 10^{35}$ erg s$^{-1}$ at the distance of Cen A), 
of which $\approx 70$ are expected to be background sources, mainly AGN (see below). 34 of these sources come from the problematic regions.
The GC sample was constructed using all the data inside $10\arcmin$ 
and contains 47 sources.

\begin{figure}
\resizebox{\hsize}{!}{\includegraphics[angle=0]{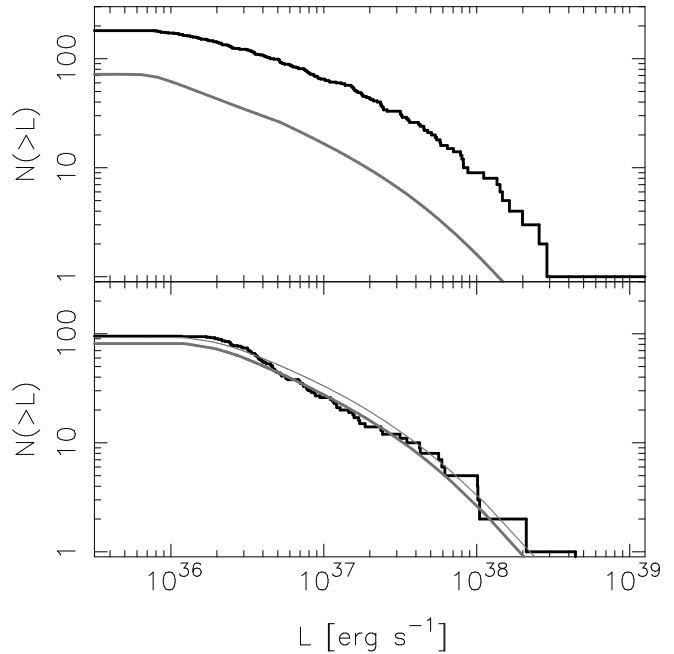}}
\caption{Upper panel: the (uncorrected) cumulative source counts 
within 5 arcmin from the center of Cen A, including both
field and GC sources (black), 
compared to the expected background counts (grey) corrected for 
incompleteness.
The sources belonging to Cen A make up the difference between the curves.
Lower panel: the cumulative source counts in the 
7.5-10.0 arcmin annulus (black) and the expected background (grey).
$\sim10$ LMXBs are expected in this annulus in addition to the background sources, 
and the thin grey line shows the expectation when these are included. The
normalization of the background curves is 1.5 times the normalization
given by \citet{Moretti}.}
\label{fig:cum}
\end{figure}

\begin{figure}
\resizebox{\hsize}{!}{\includegraphics[angle=0]{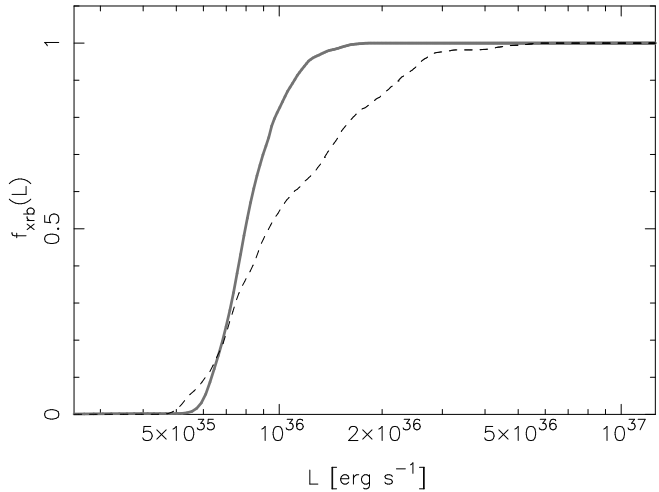}}
\caption{The incompleteness function for primordial (non-GC) X-ray binaries within 5 arcmin from the center of Cen A (solid curve) and for the GC sources within 10 arcmin (dashed line).
}
\label{fig:inc}
\end{figure}

The raw LF obtained by binning the sources over luminosity needs to be corrected for 
incompleteness and contamination
from background sources.

\subsection{Background sources}

As in \citet{Voss-cena,Voss-m31}, we used the background
source counts from  \citet{Moretti}, converting 
their soft band results into the
full 0.5-8.0 band.  Previously  \citep{Voss-cena} we found that the density of resolved CXB sources
in the direction towards  Cen A was $\sim 50\%$ higher than expected from this calculation,
whereas the shape of their $\log N-\log S$ distribution was predicted correctly.
Although this excess is larger than the variance in the CXB source density, 
$\sim 20-25\%$, which is typically quoted in the literature,
it does not contradict observations of fields where the number
of background sources
is enhanced due to nearby large-scale structures \citep{Cappelluti,Hudaverdi}.
Indeed, the Cen A field is located in the direction  of the Hydra-Centaurus
and Shapley superclusters \citep{Raychaudhury1989}, which may 
be the main reason for the observed enhancement of the background source density.
The relatively low Galactic latitude is also likely to play some role, increasing the number of foreground sources.

We therefore set the normalization of the background $\log N-\log S$ to 
1.5 times the nominal value from \citet{Moretti}. A lower (higher) value of 
the normalization would cause
the observed LF of field sources to become steeper (shallower).
In the lower panel of Figure \ref{fig:cum} we compare predicted CXB source counts with  the observed luminosity distribution of sources  detected in the 7.5-10.0 arcmin annulus.  In this region the contribution of X-ray binaries located in Cen A  is small ($\sim 10$, estimated from the density of field sources in
the inner parts of the galaxy, assuming the density of field LMXBs follow the \textit{K}-band flux) in comparison with the predicted number of CXB sources ($\sim 80$). The source luminosities were calculated assuming the distance of Cen~A; the predicted luminosity distribution for background sources was corrected for incompleteness, as described below. 
The figure demonstrates good agreement between the two distributions, both in normalization and shape. Indeed a KS-test shows that the observed distribution is compatible 
(with a KS probability of 70\%) with
a model consisting of the expected background and 10 LMXBs, which are assumed to follow
the LF of \citet{Gilfanov}. This model is represented by the thin grey line in
the lower panel of Figure \ref{fig:cum}. In agreement with \citet{Voss-cena}, we 
therefore conclude that deviations from the deep-field $\log N-\log S$ must be 
relatively modest 
\citep[this was also found for the two regions studied by][]{Hudaverdi}.
Therefore the possible errors from adopting the background 
$\log N-\log S$ of \citet{Moretti} are smaller than the statistical errors of our
sample, and they do not significantly influence our conclusions.
We note that the LF of GC sources does not need to be corrected for the contribution of background sources,
as the number of such sources coincident with GCs is negligibly small.

The upper panel of Figure \ref{fig:cum} shows the cumulative number counts of all sources
within 5 arcmin from the center of Cen A, compared to the background
expectation. The difference between the two curves arises from sources belonging
to Cen A. The vast majority of these sources are LMXBs.
Potentially a significant number of HMXBs ($\sim30$ above our detection limit) 
could be present in Cen A
\citep{Voss-cena}, but most of these are faint and in the very inner part of the galaxy,
which is excluded from our analysis, or in the dust lanes, where faint sources
are also excluded from our analysis.

\subsection{Incompleteness correction}
 
To calculate the completeness of the source samples, we used the method described
and tested in \citet{Voss-cena,Voss-m31}. In this method, the incompleteness function at a given luminosity is calculated as the fraction of pixels in which a source with this luminosity would be detected, weighted by the expected spatial  
distribution of  sources. This requires the calculation of the sensitivity for each pixel in the image.
To this end, for each of the used {\tt wavdetect} detection scales we computed the threshold
sensitivity on a grid of the positions on the image (16 azimuthal
angles, 40 radii from the center of Cen A) by inverting the detection
method. At each image position the PSF was found and
the local background levels were taken from the normalized
background maps created by {\tt wavdetect}. The sensitivity for any
given position on the image was found from interpolation of the grid
values. 

For the field sample, the spatial distribution of the LMXBs was assumed to follow the distribution of the \textit{K}-band light from the 2MASS Large Galaxy Atlas \citep{Jarret} image of Cen A, whereas the distribution of background sources was assumed flat on the scales under consideration.
In computing the incompleteness function for the GC sample we  assumed equal probability
of hosting an LMXB for all of the GCs in our catalogues. This assumption
is obviously inaccurate, as more massive/compact GCs are more likely to host LMXBs 
\citep[e.g.][]{Sarazin,Jordan2004,Sivakoff2007}.
However, if the spatial distribution of GCs does not depend strongly on their structural parameters,
this calculation gives a reasonable estimate
of the incompleteness function for GC sources.
As mentioned above, the background contamination can be neglected for GC sources.

For a given region, the differential LF $dN_{xrb}/dL$ of LMXBs is constructed 
using the following prescription: 
\begin{equation}
\frac{dN_{xrb}}{dL}=\frac{1}{f_{xrb}(L)}
\left( \frac{dN_o}{dL} -  \frac{f_b(L)}{4\pi D^2} \frac{dN_b}{dS} \right)
\end{equation}
where $dN_o/dL$ is the raw observed LF of all sources, $f_{xrb}(L)$ and  $f_b(L)$ are the incompleteness functions for X-ray binaries and CXB sources, $dN_b/dS$ is the predicted $\log(N)-\log(S)$ distribution for CXB sources, and $D$ is the distance to Cen A. The raw observed LF distribution is constructed down to the luminosity limit $L_{min}$ at which the incompleteness equals $f_{xrb}(L_{min})=0.5$. This sets up the limiting luminosity to which the final LF is constructed.
The luminosity limit  varies somewhat,
but not dramatically,
depending on the region being analyzed and the assumed source distribution,
from $L_{min}\approx 7.0\times10^{35}$ erg s$^{-1}$ for field LMXBs 
within 2.5 arcmin from the center of Cen~A, to $L_{min}\sim10^{36}$ erg s$^{-1}$
for background sources in the $5.0-7.5$ annulus. 
Further out the incompleteness quickly becomes severe, with the 0.5 
completeness limit being at 
$L_{min}\sim3\times10^{36}$ erg s$^{-1}$ in $7.5-10.0$ arcmin annulus 
(from which only GC LMXBs are included in the present study, as the number
of field LMXBs are negligible in this region). The incompleteness functions
for the GC and field source samples are shown in Fig. \ref{fig:inc}.
In the inner parts
the main factor limiting the sensitivity is the strong diffuse emission,
whereas further out it is the increased size of the PSF, and the fact that not
all regions are covered by all of the observations. The most sensitive
region is therefore at distances of about 2 arcmin from the centre, where
sources can be detected down to $\sim5\times10^{36}$ erg s$^{-1}$.

\section{Results}
\label{sect:results}

\begin{figure}
\resizebox{\hsize}{!}{\includegraphics[angle=0]{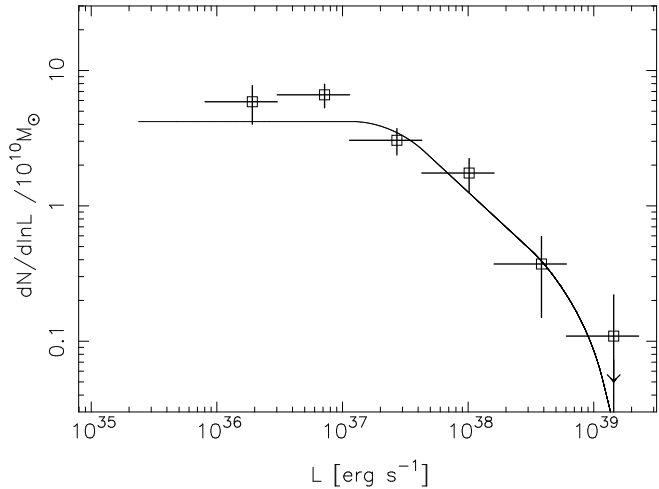}}
\caption{The LF of low-mass X-ray binaries within 5 arcmin from the center of Cen A, normalized to the stellar mass including GC sources.
The contribution of resolved background sources has been subtracted using the CXB source counts of \citet{Moretti} with normalization increased by a factor of 1.5, as described in the text. The luminosity function is corrected for incompleteness.
The solid line is the average LF of LMXBs in nearby galaxies \citep{Gilfanov}.
}
\label{fig:all}
\end{figure}

\begin{figure}
\resizebox{\hsize}{!}{\includegraphics[angle=0]{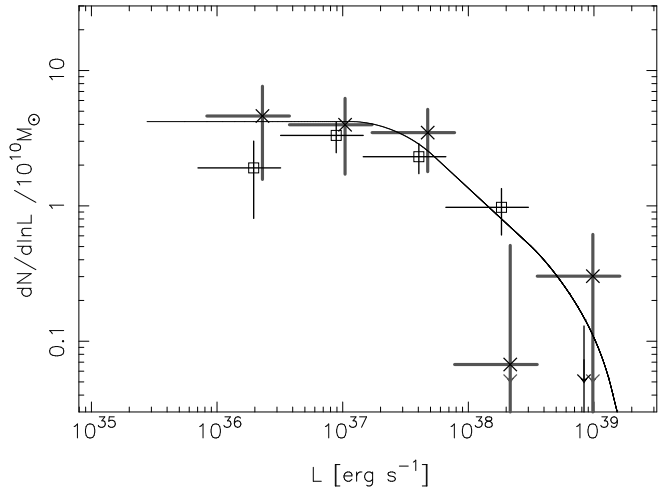}}
\caption{The LF of X-ray binaries in  the inner  $0.0-2.5$ arcmin (open squares) and $2.5-7.5$ arcmin (crosses)
annulus. The latter region extends outside the 5.0 arcmin
used for analysing field sources in the rest of this paper, to
maximally probe radial differences.
Although the background and incompleteness corrections differ strongly in the two regions, the final luminosity distributions are consistent with each other.}
\label{fig:three}
\end{figure}

In Fig. \ref{fig:all} the LF of all the LMXBs within a radius of 5
arcmin is shown. It is obvious from this figure that at luminosities
above $\sim10^{37}$ erg s$^{-1}$ the LF is consistent with a power law with the slope $\Gamma\sim 1.8-2.0$, found in previous studies of populations of compact sources in early type galaxies  \cite{Gilfanov,Kim}. It becomes nearly flat in $dN/d(\ln L)$ units below this luminosity, also in agreement with the behavior found earlier, namely  in the Milky Way, bulge of M31 and in the Cen A galaxy itself, but based on a much smaller dataset \citep{Gilfanov,Voss-cena,Voss-m31}.

We applied various statistical tests to perform a more quantitative comparison with the results of previous studies.
First we tested if the data are consistent with a single power law using the Kolmogorov-Smirnov (KS) test. 
We find that a power-law slope of $\Gamma=1.8$ is
unacceptable with a probability of $5\cdot10^{-7}$, 
whereas more shallow slopes in the range 1.3-1.4 are acceptable  with the K-S probability of $\gtrsim 10\%$.
The broken power-law fit to the data with all parameters free does not give constraining results, 
mainly due to insufficient statistics of the LF in the bright end. On the other hand, the high 
luminosity slope of the LMXB LF is rather well constrained based on the data for more massive 
(and more distant) elliptical galaxies \citep{Gilfanov,Kim}. 
We therefore fix the high luminosity index at  $\Gamma=1.8$ and fit the data, leaving the break 
luminosity and the power-law index of LF below the break free parameters of the fit. The best fit 
values of these parameters are: $L_b=(3.9^{+1.6}_{-2.3})\cdot 10^{37}$  erg/s and  $\Gamma_2=1.2\pm0.1$. 
A model with the low luminosity index fixed at $\Gamma_2=1$ is also consistent with the data with a 
KS probability of 
67\% \footnote{We note that this is not consistent with the errors on the
low luminosity slope found from the maximum likelihood fit. This is
possible because the two tests used to find the best fit and the goodness
of the fit are independent (unlike the $\chi^2$ test).}
, resulting in a best fit value of the break luminosity 
$L_b=(1.7\pm 0.7)\cdot 10^{37}$  erg/s. These numbers are consistent with the parameters from 
\citet{Gilfanov}. However, they are lower than the value of 
$5.0^{+1.0}_{-0.7}\cdot 10^{37}$ erg s$^{-1}$ found by \citet{Voss-cena}. That study was based on
shallower data, for which it was not possible to reliably exclude source contamination from the
extended emission, and therefore the results of our study are much more reliable, even if the deeper
data do not improve the statistics dramatically.

We also compare the luminosity distribution in the central $0.0-2.5$ arcmin to the
distribution in the $2.5-7.5$ arcmin annulus.
As these two regions are subject to different systematic effects, a comparison between them can be used to assess the accuracy of our analysis. 
Near the center diffuse gas and variable absorption are primary factors, while the background subtraction is unimportant. In the outer region the background subtraction is the most important effect.
Fig. \ref{fig:three} demonstrates that LFs for these two regions  are consistent with each other within statistical errors.

\begin{figure}
\resizebox{\hsize}{!}{\includegraphics[angle=0]{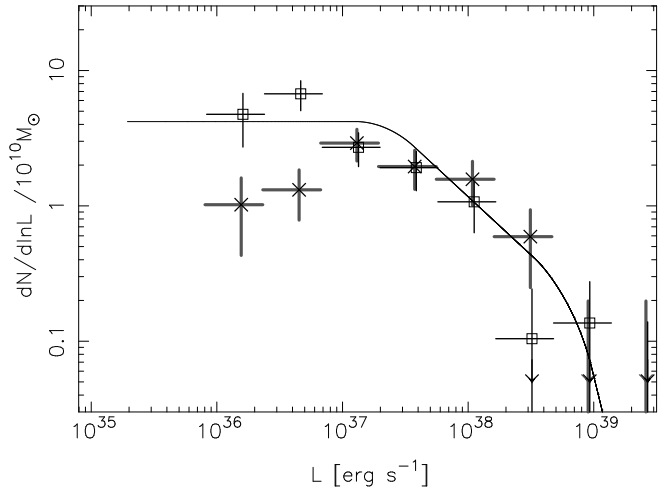}}
\caption{LFs of the field (non-GC, squares and GC (crosses, arbitrary normalization) samples. The contribution of  background sources was subtracted and incompleteness correction was applied. The relative paucity of faint sources in the LF of GC sources is apparent.
The two distributions are different at the  $\approx 3.4 \sigma$ level. Note that although there are more
sources in the field sample, the errors are comparable to
the GC sample, due to the (subtracted) contribution of CXB sources.}
\label{fig:GC}
\end{figure}

\subsection{Globular cluster and field sources}

Fig. \ref{fig:GC} compares the LFs of the GC and field ($0.0-5.0$ arcmin) samples. The two distributions appear
to be different, with the LF of the GC sample having a pronounced deficit of faint sources, $\log(L_X)\la 36.5$. The 
difference in corrections (background subtraction and incompleteness corrections)
makes  the use of the  KS-test for statistical comparison of the two distributions non-trivial. 
For this reason, we chose to employ a simpler test, in which we compare
ratios of faint to bright sources, $R=N_{faint}/N_{bright}$.
We define  faint and bright sources as those in the luminosity range
from $10^{36}$ to $10^{37}$ erg s$^{-1}$ and 
above $10^{37}$ erg s$^{-1}$ respectively. The boundary was chosen rather arbitrarily, motivated by the shape of the luminosity function and the requirement to optimize the numbers of sources in both bins for both samples in order to be sensitive to differences in the LFs. The final result is not particularly sensitive to the precise value of this boundary.

The computed faint-to-bright ratios for the field and GC samples are presented in Table \ref{tab:ratios}, where they are compared to the field and spectroscopically confirmed GCs LMXBs in
M31 from the study of \citet{Voss-m31}.
Because of the non-Gaussianity of  errors in the ratio $R$, caused by the small numbers of sources, we use Monte Carlo simulations to assess the errors and significances of
our results. In each Monte-Carlo run we  recompute  the number of sources in each luminosity bin. For each observed source a random number
is drawn from a Poisson distribution with an expectation value of
1, and the same corrections are applied to the obtained Monte-Carlo sample as the ones used to correct the observed sample.
These numbers are then summed, giving the $N_{faint}$ and $N_{bright}$ of
the Monte Carlo realization from which the value of $R$ is computed. The dispersion of the latter is used to estimate the uncertainty in $R$, cited in Table \ref{tab:ratios}.
The statistical significance of two LFs 
being different is estimated based on the null hypothesis $R_F-R_{GC}=0$,
which is calculated from
the fraction of $10^6$ Monte Carlo realizations in which $R_F-R_{GC}>0$. 
From these calculations we find $R_{GC}=0.56^{+0.20}_{-0.16}$, which is well below the value for the field sample,  $R_F=2.50^{+0.82}_{-0.62}$. The statistical significance of the difference between the two distributions 
is 99.97\% which corresponds to 3.4$\sigma$. 
As an additional check, we repeated the above analysis  with the cleanest
sample possible, entirely excluding the regions marked by boxes in Fig.\ref{fig:nice}, in addition to the jet and nucleus region. With this reduced sample, we still find that the field and GC distributions are different at the 99.89\% confidence level ($\approx 3.1\sigma$).
We are therefore confident that the population of LMXBs in GCs is different from
the field population, with a relative underabundance of faint sources in the GCs,
and that this result is not caused by contamination of our sample by spurious 
sources from the regions
with strong diffuse emission.

\section{Discussion}

\subsection{Field sources}

The luminosity function of low-mass X-ray binaries in old stellar environments is fairly well studied in the bright source limit, $\log(L_X)\ga 37.5$, where it was shown to follow a rather steep power law with slope $\Gamma\approx 1.8-1.9$ \citep{Gilfanov,Kim}. It steepens further above the Eddington limit for a neutron star, $\log(L_X)\ga 38.5-38.6$, and no LMXBs significantly more luminous than $\log(L_X)\sim 39.5$ appear to exist. This behavior seems to be universal for compact X-ray sources associated with old stellar populations. With a large number of spiral and elliptical galaxies extensively studied by different groups of researchers, no significant deviations from this behavior have been identified.
Due to obvious restrictions resulting from the sensitivity of a typical
\textit{Chandra} observation, the faint end of the 
luminosity function of point sources has not been studied as extensively
as the bright end, and so a consensus has not emerged.

With the currently available \textit{Chandra} data it is possible to study the LF of compact sources  well 
below $10^{37}$ erg s$^{-1}$ only in a handful of sufficiently large and nearby galaxies;  M31, Cen A and NGC3379 being the most significant among them. The LMXB LF is also known for the Milky Way. It has been found that LMXB LFs in the Milky Way and M31 show consistent behavior, with a clear low-luminosity break at $\log(L_X)\sim 37.5$
below which they follow a $dN/dL\propto L^{-1}$ power law \citep{Gilfanov,Voss-m31}.
Above the break, the LF shape is similar to that of compact sources in elliptical galaxies.
In a previous study of Cen A, based on $\sim$200 ks of exposure, it was shown
that the LF of compact sources there also exhibits a break at a few times $10^{37}$ erg s$^{-1}$
\citep{Voss-cena}. However, the statistical quality of the data did not permit an accurate constraint of the position of the break and the slope of the luminosity distribution. The study of two other nearby early-type galaxies gave results to the contrary,  claiming that the $\Gamma\approx1.8$ power law continues below $\log(L_X)\sim 37$ \citep{Kim2006}. It should be mentioned that the latter result was based on a more shallow dataset than the one in which the break was found. 
It remained unclear whether this discrepancy is caused by the statistical effects and/or systematic differences in the data treatment (in particular, correction for the sample incompleteness) or, on the contrary,  demonstrates that properties of the LMXB population may differ between the early and late type galaxies and even between early-type galaxies with different star-formation history. The latter possibility would not be entirely unexpected, because the population of LMXBs does evolve on $\sim$Gyr timescales.  

With the data of the very deep VLP program of Cen A observations we have  demonstrated with high confidence the presence of the low luminosity break in the LF of compact sources in this galaxy. Its overall qualitative behavior  and particular values of the best fit parameters agree well with the results for the Milky Way and M31.  

Although a complete theoretical description of the LMXB LF is yet to be created, a plausible scenario has been sketched by \citet{Postnov} (see also \citet{Bildsten}). They suggested that the break in the luminosity distribution of LMXBs corresponds to the transition from binary systems, in which the mass transfer is driven by magnetic braking of the secondary, to the binary systems losing orbital angular momentum through the emission of gravitational waves. In this interpretation the low luminosity part of the population obeying the $dN/dL\propto L^{-1}$ law is dominated by short orbital period binaries with very low mass donors, $M_d\la 0.4M_\odot$. This prediction allows, in principle, direct observational proof through study of the properties of individual binaries in the Milky Way and nearby galaxies. While this is yet to be done, the reliable determination of the LMXB shape and accurate measurements of its parameters provided by observations mentioned above and reported in this paper lend indirect support to this scenario.

\subsection{Globular cluster sources}

The difference between luminosity distributions of primordial and dynamically formed binaries was first suggested by \citet{Voss-m31} based on the analysis of LFs of LMXBs in the field, in
GCs and in the nucleus of M31. To obtain a statistically significant result they had to combine
GC sources   with LMXBs in the nucleus of M31, also shown to be formed dynamically in close stellar encounters
\citep{Voss-coll}.
Studies of populations of compact sources in NGC 3379 \citep{Fabbiano2007},
NGC 4278 and NGC4697 \citep{Kim2009} also find a relative dearth of
low luminosity LMXBs in GCs, even if it is only possible to probe the
LFs to a limit of $\sim5\times10^{36}$ erg s$^{-1}$ in these galaxies. 
Also analysis of early Cen A data 
\citep{Woodley2008} indicate a relative dearth of low luminosity LMXBs in globular
clusters.
However, the latter analysis yielded results of low 
statistical significance,
and, more importantly, ignored the issue of background source contribution. The slope of the CXB $\log N -\log S$
is $\Gamma\sim1.4-1.6$ in the flux range of interest \citep{Moretti}. As this is steeper than the
LF of the field LMXBs, to neglect their contribution may lead to artificially high
confidence estimates.  Furthermore, the study of \citet{Woodley2008}
did not take incompleteness effects into account, and they concluded that it
was not clear if their analysis was not compromised by  the sample incompleteness.
The results presented above are therefore the most robust and statistically significant evidence
yet for a difference between the luminosity distributions of LMXBs in the field and in globular clusters.

Taken at face value, the observed difference between LFs suggests that LMXBs in the field and in globular clusters belong to two different populations. This may finally falsify the hypothesis that all LMXBs in galaxies are formed in globular clusters \citep[as first suggested by][]{White}. 

It is not likely that any further significant improvement
in the statistical quality of the LFs and confidence level of this result can be achieved with yet longer observations of the Cen A galaxy, or, in general,  with other observations of individual galaxies. The observed fall-off of the GC XLF at low luminosities indicates that we are picking up the majority of the globular cluster LMXBs and no  significant further increase in their numbers should be expected. On the other hand, due to the $L^{-1}$ behavior of the luminosity distribution, the sample of field LMXBs is increasing only logarithmically with flux. Moreover, the faint end of the source sample will become increasingly polluted with the background sources, for which the flux distribution is steeper, and the increased source density may lead to increased confusion effects, especially in the inner region of the galaxy, which will further compromise the sample of faint sources.
A more realistic way to obtain high-quality XLF of the dynamically formed sources suitable for a meaningful analysis is to combine data for a number of nearby galaxies, for which \textit{Chandra} data have a sufficiently low sensitivity threshold. The same is true with respect to the luminosity function of the field sources.

\subsection{Interpretation}

The disproportionately (with respect to the stellar mass) large number of X-ray binaries found in globular clusters can be explained by the formation of binaries in close stellar encounters \citep{Clark}. 
The following three processes are believed to be the main channels of dynamical formation of X-ray binaries
in high stellar density environments typical of globular clusters: 
(i) a tidal capture of a neutron star (NS) by a non-degenerate single star \citep{Fabian}, 
(ii)  a collision of a NS with an evolved star on the subgiant or red giant branch (RGB)  \citep{Verbunt} and 
(iii) an exchange reaction, in which a NS exchanges place with a star in a pre-existing binary during a close binary-single encounter \citep{Hills}. 
With the possible exception of collisions with evolved stars which may lead to formation of an ultra-compact binary, it is difficult or  impossible to observationally distinguish between the products of these processes, and it is therefore difficult to determine their relative contributions to the populations of X-ray binaries in GCs.
Theoretical estimates show that they may be making comparable contributions \citep{Voss-coll}, but
the numbers may also depend on GC parameters  
\citep[see e.g.][]{Ivanova2008}.

Collisions between neutron stars and red (sub-) giants can lead to the formation of an X-ray binary,
in which the donor star is a white/brown dwarf or a helium star,
depending on the evolutionary stage of the evolved star before the
collision \citep{Ivanova}. This may offer  a plausible explanation for the observed deficit of
faint sources in GCs. 
As the donor star has lost all of its hydrogen envelope in the course of the common envelope phase, the material in the accretion disk will be He-rich. Because of the 4 times higher ionization potential of He, the accretion disk instability will appear at higher $\dot{M}$ value than in the case of an accretion disk of solar composition.  
Quantitatively, the critical accretion rate value $\dot{M}_{\rm crit}$ is $\sim 20$ times larger for a pure He disk than for a disk composed of solar abundance material \citep{Lasota}. The theoretical lower limit on the bolometric luminosity of persistent systems with orbital period of $P_{orb}=60$ min (approximately the shortest period possible for non-degenerate donor stars) changes from $\sim 10^{36}$ erg/s for a solar abundance disk to $\sim 2\cdot 10^{37}$ erg/s for a pure He disk \citep[assuming accretion efficiency of 0.2]{Lasota}. It is smaller for more compact systems, decreasing as $P_{orb}^{\approx 1.6-1.7}$. We note that due to the limited energy
range covered by \textit{Chandra}, the observational luminosities are somewhat smaller than the theoretical estimates, and corrections due to this suffer from relatively large
spectral uncertainties.

The majority of field (non-GC) LMXBs are likely to have main sequence donors, unless they are formed in globular 
clusters. This is expected theoretically in the models of primordial binary evolution and is also confirmed 
by the statistics of X-ray binaries in the Milky Way \citep{Liu}.
If a significant fraction of X-ray binaries in globular clusters have He donors, this may explain the smaller fraction 
of low luminosity systems as compared with the field population, as they will not be observed as persistent sources below 
the transience limit $L_{crit}$. In this scenario not only is the number of faint sources reduced, but also the 
number of luminous ones is possibly increased due to intrinsically faint systems that are currently in outburst. 
Calculations of the luminosity distribution of X-ray binaries accounting for these 
effects appear to be beyond the predictive power of the current generation of population synthesis codes.
The problem is further complicated by the fact that the accurate calculation of the formation rates of various types of 
binaries is insufficient, as   evolution, X-ray lifetimes and transient behavior of the binaries should also be considered. 
Estimates of \cite{Voss-coll} indicate that He-accreting systems may contribute about $\sim 1/3$ to the LMXB population of 
globular clusters, in rough agreement with the statistics of globular cluster sources in the Milky Way. This may be 
sufficient to explain the fall-off of the GC LF to low luminosities, although modelling would  
depend on the details of the luminosity distribution of X-ray binaries formed via other formation mechanisms (which 
may differ from the LF of field sources) and on the fraction of He-accreting systems in outburst.  On the other hand, 
the fact that the value of the critical luminosity limit for persistent He-accreting systems, $\sim 10^{37}$ erg/s,  
is in the range where the GC LF fall-off begins,  gives an indirect support to the proposed scenario. A more direct 
observational test may come from statistics of transient sources in globular clusters in nearby galaxies.

\section{Conclusions}
We have studied the luminosity function of LMXBs in the early-type galaxy
Centaurus A (NGC 5128),  
with particular emphasis on its behavior in the low luminosity regime  
and special attention to various systematic and instrumental effects  
that may compromise the faint end  of the luminosity distribution of  
compact sources in such a complicated galaxy.
We demonstrated that sub- 
structures in the diffuse emission can significantly pollute the  
source list produced  by the automated point source detection  
software. This effect, less obvious with previous shorter  
observations, became especially evident with the deep $\sim$800 ks  
exposure collected by the \textit{Chandra} VLP program for this galaxy.
We also performed accurate corrections for incompleteness and removal of  
the  contribution of resolved background sources.

We confirm with high statistical confidence that the luminosity  
function of the LMXBs flattens significantly below $\log(L)\sim 37.5$  
and is inconsistent with extrapolation of the $\Gamma\approx 1.8$ power  
law observed in numerous elliptical galaxies at higher luminosities.  
In the low luminosity regime it is consistent with the logarithmically  
flat distribution predicted by some theories of binary evolution and  
previously observed in two other nearby galaxies

We find with a confidence of $99.97\%$ that the luminosity  
distribution of LMXBs in globular clusters differs from that of the  
field (non-globular cluster) LMXBs, with a relative lack of 
low-luminosity sources in the globular cluster sample. This may finally  
falsify the hypothesis that the entire population of LMXBs in early  
type galaxies has been formed in globular clusters.
As a plausible explanation for this difference we suggest that there is
a large  
fraction of binaries with He-rich donors in the LMXB population in  
globular clusters. These systems will show transient behavior at $\sim  
20$ times higher accretion rate than X-ray binaries with a normal main  
sequence donor and therefore would not be detected among persistent  
sources with $\log(L_X)\la 37.0-37.5$. In globular clusters, they may  
be created in sufficient numbers  in collisions of compact objects  
with red (sub-) giants, whereas their number among the field sources  
is small. An observational check for this scenario may come from statistics  
of transient sources in globular clusters.

%% If you wish to include an acknowledgments section in your paper,
%% separate it off from the body of the text using the \acknowledgments
%% command.

%% Included in this acknowledgments section are examples of the
%% AASTeX hypertext markup commands. Use \url without the optional [HREF]
%% argument when you want to print the url directly in the text. Otherwise,
%% use either \url or \anchor, with the HREF as the first argument and the
%% text to be printed in the second.

\acknowledgments
This research has made use of data obtained from the \textit{Chandra} Data Archive 
and software provided by the \textit{Chandra} X-ray Center (CXC) in the application 
package CIAO. This work was supported by the NASA grants GO7-8105X and
NAS8-03060. R.V. acknowledges the support by the DFG cluster of excellence 'Origin and
Structure of the Universe' (http://www.universe-cluster.de). CLS and GRS
were supported in part by Hubble Grants HST-GO-10597.03-A, HST-GO-10582.02-A,
and HST-GO-10835.01-A, and Chandra Grant GO8-9085X. MSH thanks the Royal
Society for support. The authors would like to thank the anonymous 
referee for constructive criticism 
of the original manuscript which helped to improve the paper.

%% To help institutions obtain information on the effectiveness of their
%% telescopes, the AAS Journals has created a group of keywords for telescope
%% facilities. A common set of keywords will make these types of searches
%% significantly easier and more accurate. In addition, they will also be
%% useful in linking papers together which utilize the same telescopes
%% within the framework of the National Virtual Observatory.
%% See the AASTeX Web site at http://www.journals.uchicago.edu/AAS/AASTeX
%% for information on obtaining the facility keywords.

%% After the acknowledgments section, use the following syntax and the
%% \facility{} macro to list the keywords of facilities used in the research
%% for the paper.  Each keyword will be checked against the master list during
%% copy editing.  Individual instruments or configurations can be provided 
%% in parentheses, after the keyword, but they will not be verified.

{\it Facilities:} \facility{\textit{Chandra} (ACIS)}.

\clearpage

\begin{deluxetable}{lcccccc}
\tabletypesize{\scriptsize}
\tablecaption{The \textit{Chandra} observations used in this paper.}
\tablewidth{0pt}
\tablehead{
\colhead{Obs-ID} & \colhead{Date} & \colhead{Instrument} & \colhead{Exp. Time} & \colhead{R.A.} & \colhead{Dec.} & \colhead{Data Mode}
}
\startdata
0316 & 1999 Dec 05 & ACIS-I & 36.18 ks & 13 25 27.61 & $-$43 01 08.90 & FAINT \\
0962 & 2000 May 17 & ACIS-I & 36.97 ks & 13 25 27.61 & $-$43 01 08.90 & FAINT \\
2987 & 2002 Sep 03 & ACIS-S & 45.18 ks & 13 25 28.69 & $-$43 00 59.70 & FAINT \\
3965 & 2003 Sep 14 & ACIS-S & 50.17 ks & 13 25 28.70 & $-$43 00 59.70 & FAINT \\
7797 & 2007 Mar 22 & ACIS-I & 98.17 ks & 13 25 19.15 & $-$43 02 42.40 & FAINT \\
7798 & 2007 Mar 27 & ACIS-I & 92.04 ks & 13 25 51.80 & $-$43 00 04.43 & FAINT \\
7799 & 2007 Mar 30 & ACIS-I & 96.04 ks & 13 25 51.80 & $-$43 00 04.43 & FAINT \\
7800 & 2007 Apr 17 & ACIS-I & 92.05 ks & 13 25 46.00 & $-$42 58 14.58 & FAINT \\
8489 & 2007 May 08 & ACIS-I & 95.18 ks & 13 25 32.79 & $-$43 01 35.13 & FAINT \\
8490 & 2007 May 30 & ACIS-I & 95.68 ks & 13 25 18.79 & $-$43 03 01.72 & FAINT \\
 
\enddata
%% Text for table notes should follow after the \enddata but before
%% the \end{deluxetable}. Make sure there is at least one \tablenotemark
%% in the table for each \tablenotetext.
\label{tab:obs}
\end{deluxetable}
\clearpage
\begin{deluxetable}{lccccccc}
\tabletypesize{\scriptsize}
\tablecaption{The ratios of faint to bright sources in Cen~A and M~31.}
\tablewidth{0pt}
\tablehead{
\colhead{Region} & \colhead{$N_{faint}$\tablenotemark{a}} & \colhead{$N_{bright}$\tablenotemark{a}} & \colhead{$B_{faint}$\tablenotemark{b}} & \colhead{$B_{bright}$\tablenotemark{b}} & \colhead{$N_{cor,faint}$\tablenotemark{c}} & \colhead{$N_{cor,bright}$\tablenotemark{c}} & \colhead{Ratio}
}
\startdata
GCs & 17 & 30 & 0.9 & 0.4 & 16.7 & 29.6 & $0.56^{+0.20}_{-0.16}$\\
Field & 96 & 50  & 43.0 & 14.6 & 88.5 & 35.4 & $2.50^{+0.82}_{-0.62}$\\
M 31 GC & 3 & 8 & 0 & 0 & 3 & 8 & $0.38^{+0.34}_{-0.24}$\\
M 31 field & 40 & 19 & 15.5 & 0.8 & 24.6 & 18.2 & $1.35^{+0.58}_{-0.42}$\\
\enddata
\tablecomments{The subscripts $faint$ and $bright$ corresponds to sources in
the $10^{36}-10^{37}$ erg s$^{-1}$ and $>10^{37}$ erg s$^{-1}$ ranges,
respectively.}
\tablenotetext{a}{Observed uncorrected number of sources.}
\tablenotetext{b}{Expected number of background sources.}
\tablenotetext{c}{Background subtracted and incompleteness corrected
number of sources (including correction for the change in survey area
below $5\times10^{36}$ erg s$^{-1}$ due to the exclusion of problematic
regions).}
\label{tab:ratios}
\end{deluxetable}

\end{document}